\documentclass[aps,prb,twocolumn, floatfix ,superscriptaddress,showpacs]{revtex4}

\usepackage{color}
\usepackage{bm}
\usepackage{amssymb}
\usepackage{amsmath}
\usepackage{amstext}
\usepackage{graphics}
\usepackage{epsfig,epstopdf}
\usepackage{placeins}
\usepackage{psfrag}
\usepackage{subfigure}
\def\ie{\emph{i.e.},\ }
\def\eg{\emph{e.g.}\ }
\def\ea{\emph{et~al.}}

\begin{document}

\graphicspath{{Figures/}}

\title{Critical theory of the topological quantum phase transition in the AKLT--SZH chain}
\author{Hong-Chen Jiang}
\affiliation{Microsoft Research, Station Q, University of
California, Santa Barbara, CA 93106}

\author{Stephan Rachel}
\affiliation{Department of Physics, Yale University, New Haven,
CT 06520}

\author{Zheng-Yu Weng}
\affiliation{Institute for Advanced Study, Tsinghua University,
Beijing, 100084, P.\,R.\,China}

\author{Shou-Cheng Zhang}
\affiliation{Department of Physics, McCullough Building,
Stanford University, Stanford, CA 94305}

\author{Zhenghan Wang}
\affiliation{Microsoft Research, Station Q, University of
California, Santa Barbara, CA 93106}


\begin{abstract}
We systematically study the phase diagram of $S=2$ spin chain by
means of density-matrix renormalization group and exact
diagonalization. 
We confirm the presence of a dimer phase in
the AKLT--SZH model and find that the whole phase boundary between
dimer and SZH phases, including the multicritical point, is a
critical line with 
central charge $c=5/2$. Finally, we
propose and confirm that this line corresponds to
$\rm SO(5)_1$ Wess--Zumino--Witten conformal field theory.
\end{abstract}

\maketitle

In recent years, investigations of topological phases and phase
transitions has attracted great attention in condensed matter
physics\cite{Qi2010} and quantum information
theory\cite{Kitaev2006}. Topological phases are characterized
by a bulk gap separating excitations from the ground state
and by the presence of gapless edge modes. A topological phase cannot be deformed
continuously into a conventional, topologically trivial phase
without going through a phase transition, in which the gap closes
and the edge modes merge with the bulk. The quantum Hall
state\cite{Klitzing1980} and the recently discovered topological
insulators\cite{Qi2010}
are examples of topological states of quantum matter.

Topological phases even appear in one dimensional (1D) systems. The Haldane phase
\cite{Haldane1983} of integer quantum spin chains is an example of a
{\it symmetry protected topological phase}\cite{Gu2009,Pollmann2010}. In 1987, Affleck, Kennedy, Lieb,
and Tasaki (AKLT)\cite{Affleck1987} introduced a family of exactly
solvable integer spin models with valence bond solid (VBS) ground
state and proved that VBS states share the key features of
Haldane gap liquids for integer spin chains. 
The ground state of AKLT model can be exactly formulated in terms of
Schwinger bosons\cite{Haldane1988},
$|\Psi ^{\rm AKLT}\rangle=\prod_{\langle ij\rangle}(a_{i}^{\dagger
}b_{j}^{\dagger }-b_{i}^{\dagger }a_{j}^{\dagger })^{S}|0\rangle,
$
where $S$ is the local spin. 
The VBS states are unique ground states of the AKLT Hamiltonian for
periodic boundary condition (PBC); for open boundary conditions
(OBC), however, the ground state is $(S+1)^2$--fold degenerate with
edge spins $S/2$. 
The parent Hamiltonian for the VBS states $|\Psi ^{\rm AKLT}\rangle$,
here for $S=2$, is conveniently defined in terms of projection operators\cite{Affleck1987},
\begin{equation}
H^{\rm AKLT}=\sum_{\langle
ij\rangle}K_3P_3(i,j)+K_4P_4(i,j),
\label{Eq:AKLT_Ham}
\end{equation}
where $K_{3},K_4>0$ and $P_3$  and $P_4$ project onto the spin 3
and spin 4 subspaces at the bond $(i,j)$, respectively.
The edge spin is boson-like $S=1$ and the square of
the time reversal operator $T$ satisfies $T^2=1$.

In 1998, Scalapino, Zhang, and Hanke (SZH) introduced an SO(5)
symmetric superspin model\cite{Scalapino1998} with an exact VBS
ground state. The local spin in the SZH model transform under
the five-dimensional vector representation of SO(5). SZH presented
an exact ground state wave function expressed as a matrix product
state of Dirac $\Gamma$ matrices, and showed that the ground state is
16-fold degenerate for OBC.
This model can be naturally mapped onto a spin 2 chain, with
fermion-like $S=3/2$ edge spin\cite{Tu2008} and $T^2=-1$.
Following SZH's work, more VBS states with higher symmetry
groups have been constructed\cite{otherVBS,Zheng2010}.

%
The ground state of SZH model is given by
$
|\Psi^{\rm SZH}\rangle= 
\sum_{\{m_i\}}
\rm{Tr}(\Gamma^{m_1}\Gamma^{m_2}\ldots\Gamma^{m_N})|m_1\cdots
m_N\rangle\ ,
$
where the $\Gamma^m$ fulfill $\Gamma^a\Gamma^b=2\delta^{ab}+2i\Gamma^{ab}$ and $m_i$ is a vector
label of the SO(5) group, which can also be interpreted as the
$m_i=-2,-1,0,1,2$ quantum numbers of spin 2. The
corresponding 
Hamiltonian is given by%
\begin{equation}
H^{\rm SZH}=\sum_{\langle
ij\rangle}J_2P_2(i,j)+J_4P_4(i,j),
\label{Eq:SZH_Ham}
\end{equation}
where $J_2, J_4>0$ and $P_2$ and $P_4$
again are the SU(2) bond projection operators onto spin 2 and 4, respectively.
The ground state is unchanged up to an SO(5) rotation.

Like for topological insulators, the bulk topology is related to the edge states of an
open chain. Consequently, AKLT and SZH models describe different topological phases.
As recently pointed out\cite{Pollmann2009}, odd integer
spin AKLT models are protected by a couple of symmetries, \eg time
reversal, while even integer spin AKLT models are solely protected
by global SU(2) symmetry. Since the first case is characterized by
half-integer edge spin, the SZH model is similar to odd spin AKLT
models (notice that the spin 3 AKLT model exhibits an edge spin 3/2
like SZH). Both, the AKLT and SZH phases are thus protected by
symmetries, but the SZH phase seems to be much more robust. Given
the topological distinction of the two ground states, we construct a
model Hamiltonian interpolating between the AKLT and SZH models:
\begin{eqnarray}
H(\alpha) &=&(1-\alpha )H^{\rm AKLT}+\alpha H^{\rm SZH} \label{Eq:AKLT_SZH_Ham}
\\[5pt]
&=&\sum_{\langle ij\rangle} \Big[\alpha P_2(i,j) +
(1-\alpha)P_3(i,j) + \beta P_4(i,j)\Big].\nonumber
\end{eqnarray}
Here, we set $J_2=K_3=1$, $J_4=K_4=\beta$. As the edge state is
robust unless the gap closes, there must exist one or several
topological quantum phase transitions (TQPT)
when varying $\alpha$ from 0 to 1.

\begin{figure}[tbp]
\centering

\vspace{0pt}
    \includegraphics[scale=1.]{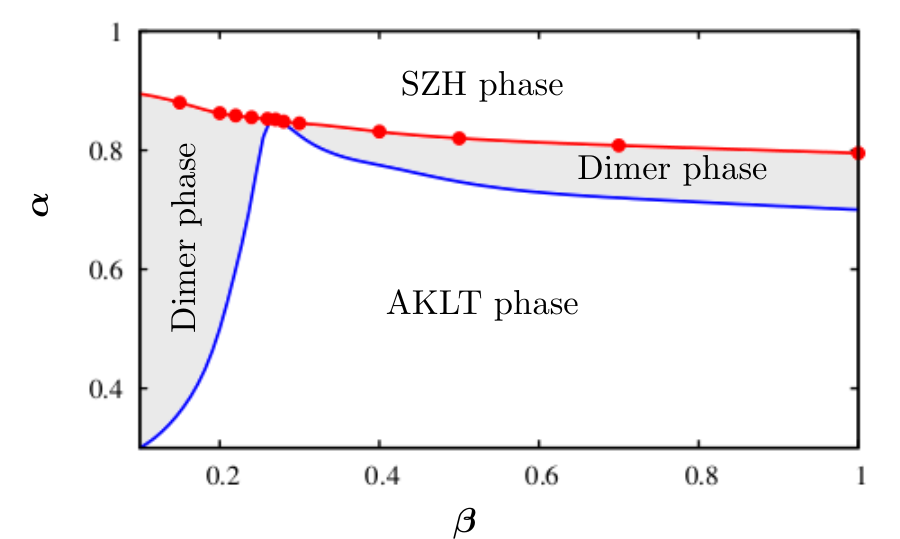}
\caption{(color online) Ground state phase diagram of the
Hamiltonian\,\eqref{Eq:AKLT_SZH_Ham} obtained by DMRG with $N = 600$ sites.
The location of second order
phase transition 
is indicated by the red line. The region above the red line belongs
to SZH phase, while the region below the dashed blue line belongs to
AKLT phase. The intermediate regime between the topologically different
SZH and AKLT phases are dimer phases.} \label{Fig:Phasediagram}
\end{figure}

In Ref.\,\onlinecite{Zang2010}, the ground state phase diagram of
the AKLT--SZH model (\ref{Eq:AKLT_SZH_Ham}) has been studied by
means of density-matrix renormalization group
(DMRG)\cite{White1992}. Between the AKLT and SZH phases, a possible
dimer phase has been found. The most interesting feature is the
presence of a multicritical (MC) point, at which a direct TQPT
occurs between the two distinct phases. However, there are still some
important questions. (1) Does the dimer phase
indeed exist in the thermodynamic limit? (2) What is the phase
diagram for $\beta<0.27$? (3) What is the central
charge and low-energy effective theory at the MC point? (4) Does the effective
theory spread over the whole phase boundary?

Motivated by these questions, in this paper we will revisit the
model \eqref{Eq:AKLT_SZH_Ham} and try to answer all open questions.
%
%
For the present study, we keep up to $m=3400$ DMRG states with
more than $16$ sweeps to get converged results,
and the truncation error is less than $10^{-9}$. 
We use both OBC and PBC with system sizes up to $N=600$ sites.

\begin{figure}[tbp]
\centering

\vspace{-10pt}
    \includegraphics[height=2.8in,width=3.6in]{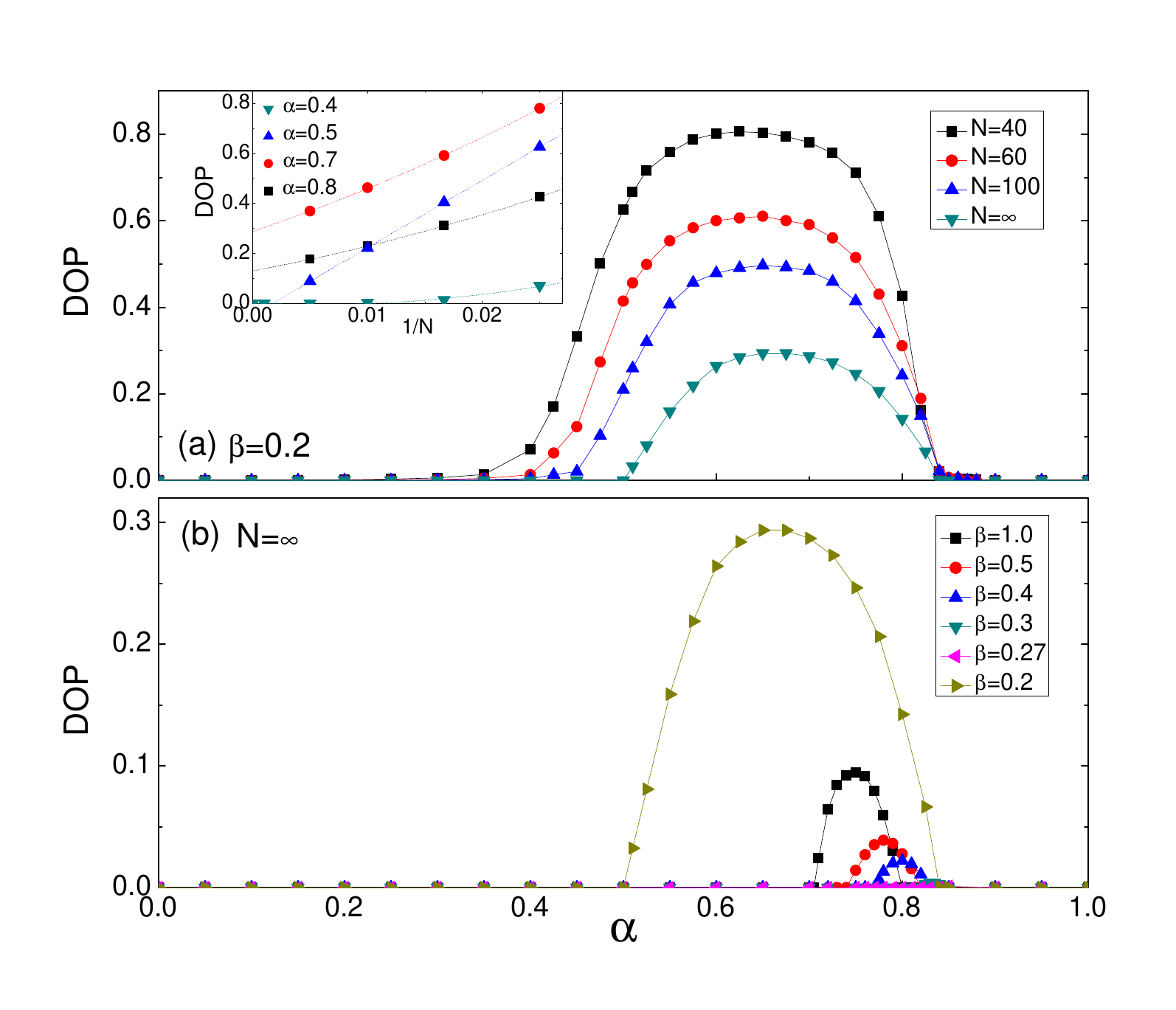}
\caption{(color online) Dimer order parameters (DOP) of the
Hamiltonian (\ref{Eq:AKLT_SZH_Ham}) with different of $\alpha$ and
$\beta$. (a) DOP for $\beta=0.2$ at system size $N=40$, $60$ and
$100$, as well as in the thermodynamic limit $N=\infty$. The
finite-size scaling is given in the inset. (b) DOP for different
$\beta$, which disappears at the multi-critical point $\beta =
0.27$.} \label{Fig:DimerOrderParameter}
\end{figure}

{\it Dimer order parameter.\,} We confirm the presence of the dimer phase for $\beta>0.27$, see the phase
diagram Fig.\,\ref{Fig:Phasediagram}. In addition, we find a much larger
dimer phase for $\beta<0.27$. To clarify the presence of the dimer
phase, we have calculated the dimer order parameter (DOP),  $\mathcal{D}=
|\langle \mathbf{S}_i\mathbf{S}_{i+1}\rangle - \langle
\mathbf{S}_{i+1}\mathbf{S}_{i+2}\rangle|\,, $ as a function of
$\alpha$ and $\beta$, where $i=N/2$ and we imposed OBC.
In Fig.\,\ref{Fig:DimerOrderParameter}(a),
exemplarily the DOP as a function of $\alpha$ at $\beta=0.2$ for
different system sizes is shown. In the inset, the finite-size
scaling is also performed to obtain the DOP in the thermodynamic limit.
As shown in
Figs.\,\ref{Fig:Phasediagram} and \ref{Fig:DimerOrderParameter},
when $\beta$ decreases from $\beta=1$,
also the maximum amplitude of DOP decreases, and the dimer phase
shrinks with $\alpha$ monotonously. Finally, at the MC point
$\beta=0.27$, the DOP vanishes for any $\alpha$. Surprisingly, for
smaller values of $\beta$ there is a revival of the dimer phase,
which becomes even larger and the magnitude of DOP increases,
indicating a very robust dimer phase at small $\beta$ region.
In the limit $\beta=0$, the $P_4$ term vanishes and we expect a huge ground
state degeneracy. Therefore, it will be extremely difficult to get
converged DMRG results and we only consider $\beta \geq 0.1$ in the
phase diagram Fig.\,\ref{Fig:Phasediagram}. Like for the general
spin 1 chain, we expect a {\it generic} dimer phase containing the
Hamiltonian $H_{\rm dimer}=\sum_i -P_0(i,i+1)$ with an enhanced
SU(5) symmetry. We performed DMRG calculations to verify this guess.
We also studied the two Hamiltonians interpolating between $H_{\rm
dimer}$ and the dimer phase for $\beta>0.27$ and for $\beta<0.27$,
respectively. In both cases, we found in the second derivative of
the ground state energy with respect to the interpolation parameter
a signal for a second order phase transition. So it might be that
the general spin 2 chain contains three different dimer phases.
The nature of the phase transition between dimer and AKLT phases seems to be a higher order (\ie third or more) phase transition. Therefore, it will be
extremely difficult to locate the exact phase boundary numerically.

\begin{figure}[tbp]
\centerline{
    \includegraphics[height=2.6in,width=3.6in]{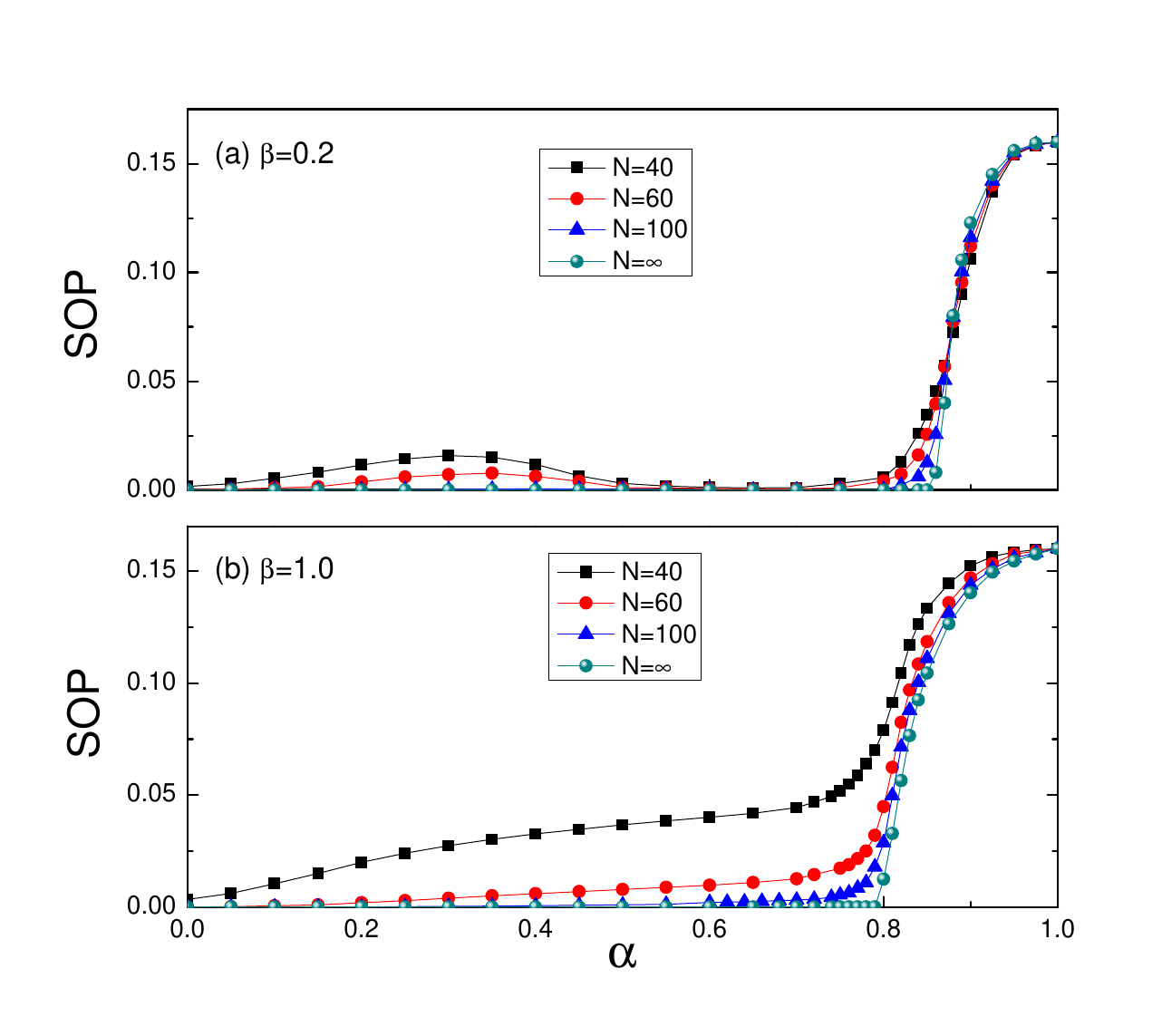}
    }
\caption{(color online) String order parameters of $H(\alpha)$
for (a) $\beta=0.2$ and (b) $\beta=1.0$, for system size
$N=40$, $60$ and $100$, and the extrapolated value for
$N=\infty$.} \label{Fig:SOP}
\end{figure}


{\it String order parameter.\,} To distinguish different topological
phases, one way is to use the hidden string order parameter (SOP)
defined as $\mathcal{G}(\theta)=\langle\hat{A}_i \exp{( i \theta
{\sum_{k=i}^{j-1}} \hat{A}_k} ) \hat{A}_j\rangle$. For spin-$S$ AKLT
model, the peak of SOP is at $\theta_c=\pi/S$ for $\hat{A}_i=S^z_i$.
But this SOP cannot be used to distinguish the AKLT and
SZH phase since it is finite in both phases. Fortunately, Tu \ea
\cite{Tu2008} proposed a SOP with $\hat{A}_i = \frac{1}{6}
(S^z_i)[(S^z_i)^2-1]$ and $\theta_c=\pi$ which {\it knows} about SO(5)
symmetry of the SZH model and which is finite only in the SZH phase.
Therefore, we can use the SOP to
characterize the TQPT between SZH and other phases (in addition to $d^2E/d\alpha^2$).
In Fig.\,\ref{Fig:SOP}, the SOP for $\beta=0.2$ and $\beta=1.0$ is
shown for several system sizes and for $N\to\infty$.
In both
the AKLT and dimer phase, $\mathcal{G}(\pi)\to 0$. In the SZH phase,
however, $\mathcal{G}(\pi)$ remains finite, and approaches its maximum value\cite{Tu2008}
$4/5^2=0.16$ in the SZH limit $\alpha=1$.

\begin{figure}[t!]
\centering
   \includegraphics[height=1.9in,width=3.6in]{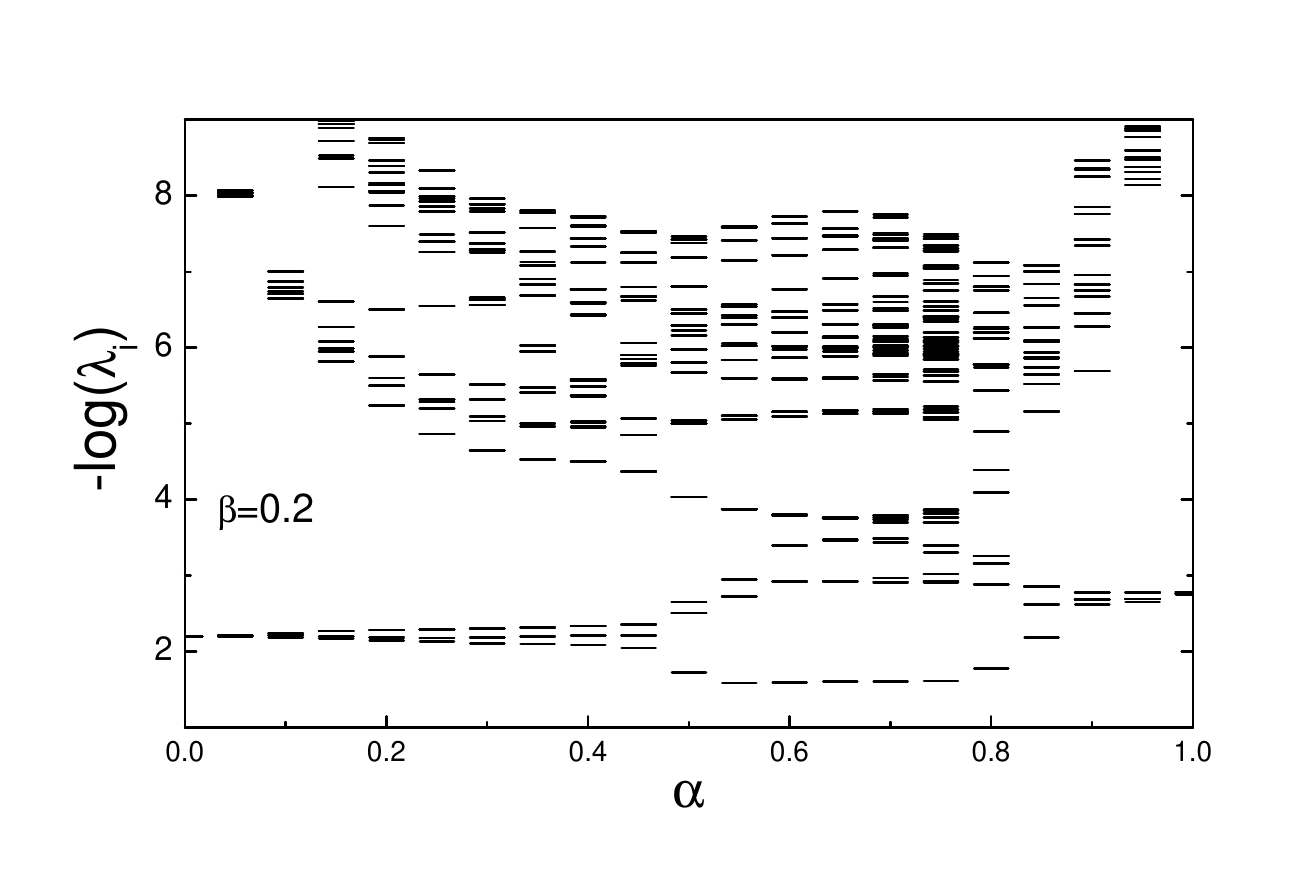}\\[-10pt]
\caption{The 100 lowest levels of the
entanglement spectrum $-\log{\lambda_i}$ for system size $N=100$ at
$\beta=0.2$. $\lambda_i$ are the eigenvalues of the reduced density
matrix.} \label{Fig:EntanglementSpectrum}
\end{figure}

{\it Entanglement spectrum.\,} To identify different phases, we can
also use the entanglement spectrum (ES), which was originally
introduced by Haldane\cite{Haldane2008} in the context of FQHE and
later applied to spin chains in a momentum basis\cite{Thomale2010}.
The ES is defined as the set of eigenvalues of the reduced density
matrix $\rho_A\equiv \rm{Tr}_B |\Psi\rangle \langle\Psi|$, with $A$
being a subsystem and $B$ the remainder of the system. Here we use
the real space ES with PBC to characterize
the different phases. As shown in
Fig.\,\ref{Fig:EntanglementSpectrum}, the lowest level of ES in AKLT
phase is 9-fold degenerate, this degeneracy is associated with the
edge spin 1. By increasing $\alpha$, a finite size splitting of the
9 ground states occurs (chain length $N=100$). After entering the
dimer phase, the lowest value of the ES becomes non-degenerate. By
further increasing $\alpha$, the system enters the SZH phase, the
lowest level of the ES becomes degenerate again, but now the
degeneracy is 16-fold in agreement with edge spin 3/2. In all the
three phases, the lowest level of ES is separated by a large gap
from the other levels.

{\it Central charge.\,} The most interesting feature is the presence
of the MC point, as shown in Fig.\,\ref{Fig:Phasediagram}. To
characterize the critical theory we should determine central charge
and scaling dimensions. The central charge can be obtained by
calculating the von Neumann entropy of a subsystem $A$ with length
$x$, defined as $S^{\rm vN} = -\rm{Tr} (\rho_A \rm{ln}\rho_A)$. For
critical systems, it has been established\cite{Calabrese2004} that
$S^{\rm vN}=(c/3)\ln(x^\prime)+\tilde c_1$ for PBC, and
$S^{\rm vN}=(c/6)\ln(2x^\prime)+\ln(g)+\tilde c_1/2$ for OBC, where $c$
is the central charge of the conformal field theory (CFT), $\tilde
c_1$ is a model dependent constant, and $g$ is Affleck and Ludwig's
universal boundary term\cite{Affleck1991b}. For finite chains, we
can use the conformal mapping $x\rightarrow x^\prime =
(N/\pi)\sin(\pi x/N)$. Using this formula the central charge
can be extracted in excellent agreement with the CFT
prediciton\cite{Stephan2008,Gils2009}.


\begin{figure}[t!]
\centerline{
    \includegraphics[height=3.0in,width=3.8in]{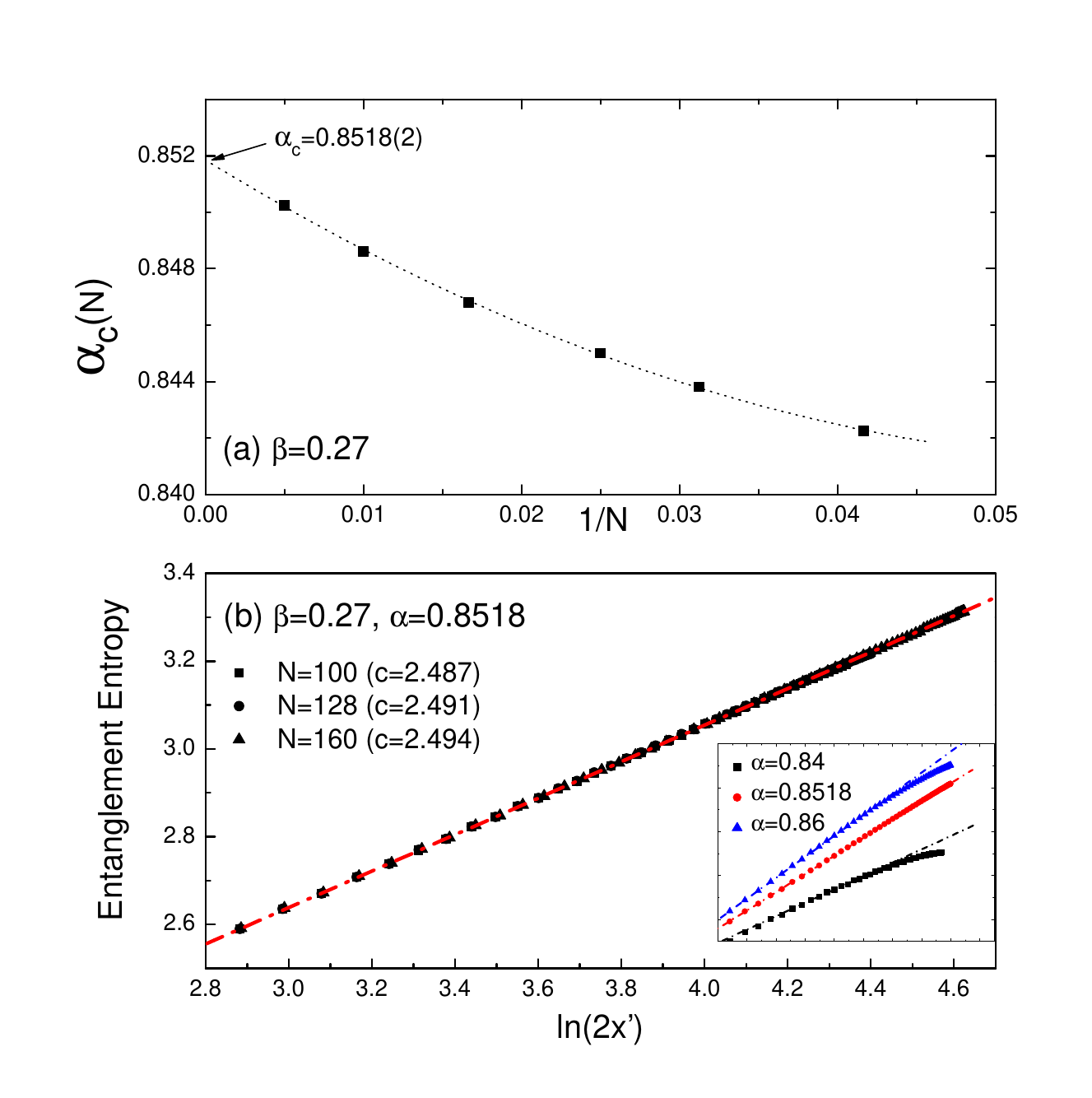}
    }
\caption{(color online) (a) Finite-size scaling of the peak position
$\alpha_c(N)$
of 
$d^2E/d\alpha^2$ as a function of $\alpha$ at $\beta=0.27$. (b) Von
Neumann entropy $S^{\rm vN}$ at the multi-critical point for
different chain length with the fitted central charge $c$. Inset:
$S^{\rm vN}$ for different parameter points is shown for
comparison.} \label{Fig:EEntropy}
\end{figure}

By performing finite-size scaling
of the peak position of $d^2E/d\alpha^2$ as a function of $\beta$ and $\alpha$,
we find the exact position of the MC point at $(\beta_c,
\alpha_c)=(0.27, 0.8518)$.
In Fig.\,\ref{Fig:EEntropy}(a), the
finite-size scaling is shown for $\beta=0.27$. In
Fig.\,\ref{Fig:EEntropy}(b), we show $S^{\rm vN}$ at the
MC point for different system sizes. The regression fit for
$S^{\rm vN}$ shows very good convergence with system size,
indicating a central charge $c=5/2$. For comparison, in the inset
of Fig.\,\ref{Fig:EEntropy}(b), we show the results for some points
away from the MC point where $S^{\rm vN}$ starts to saturate indicating the
opening of a gap. In the same way, we find  $c=5/2$ for the whole critical line
between SZH and dimer phases.
%

{\it Effective field theory.\,}
To find the corresponding CFT, we have listed all the simple
corresponding candidates with central charge $c=5/2$ in
Table\,\ref{Table:CFT_Candidate}. $S_2/S_1$ is the ratio between the
scaling dimensions of the second and first non--trivial primary
fields. To characterize the CFT theory at the critical line, we
rescale and match the lowest three finite-size energy levels
obtained numerically by exact diagonalization (ED) for systems with
up to $L = 12$ sites to the form of the spectrum of a
CFT\cite{CardyCFT},
\begin{eqnarray}
E(L) = E_1 L + {\frac{2\pi v}{L}} \left(-\frac{c}{12} + h + \bar{h}\right)\ .
\end{eqnarray}
Here the velocity $v$ is an overall non--universal scale factor and
the scaling dimensions $h+\bar{h}$ take the form $h = h^0+n,\bar{h}
=\bar{h}^0+\bar{n}$, with $n$ and $\bar{n}$ non-negative integers.
$h^0$ and $\bar{h}^0$ are the holomorphic and antiholomorphic
conformal weights of primary fields in the CFT. The momenta (in
units $2\pi/L$) are such that $k = h-\bar{h}$ or $k = h-\bar{h}
+L/2$. We consider the energy spectrum at $k=0$ with scaling
dimension $S=2h$ (\ie $h=\bar{h}$). We find that $S_2/S_1=1.56$ at
the critical line. Comparing this value with the values in
Table\,\ref{Table:CFT_Candidate}, we find that only $\rm SO(5)_1$ is
compatible with our data, while all the other candidates can be
explicitly ruled out. To further confirm this, we also calculate the
ratio $(1+S_2)/(1+S_1)$ at $k=2\pi/L$. We find $(1+S_2)/(1+S_1) =
1.27$, which is also consistent with the value $(1+1)/(1+5/8)=1.23$
(here $n=1$) of $\rm SO(5)_1$ CFT, by considering the finite-size
effect. We conclude that the critical line is described by $\rm
SO(5)_1$ Wess--Zumino--Witten (WZW) model.

\begin{table}[t]
\caption{List of CFT with central charge $c=5/2$. $S_2/S_1$ is the
ratio of the second and first primary scaling dimensions.
}
\label{Table:CFT_Candidate}\centering \vspace{2mm}
\begin{tabular}[b]{|c|c|c|c|c|c|}
\hline \hspace{0mm}&\hspace{0mm} $\rm SO(5)_1$\hspace{0mm} &
\hspace{0mm} $\rm SU(2)_{10} $\hspace{0mm} & \hspace{0mm} $\rm
SU(2)_2\times
U(1)$\hspace{0mm} & $\rm SU(2)_4\times Ising$ & MC \hspace{0mm}\\
\hline$\frac{S_2}{S_1}$&$1.6$ & 2.67 & 2.67 & 2.67 & 1.56 \\ \hline
\end{tabular}
\end{table}

This result is particularly interesting since we expect three
integrable models in the phase diagram of the general spin 2
chain\cite{Kennedy1992b}: the Takhtajan--Babujian chain (SU(2)$_4$
WZW with $c=2$), the permutation operator (SU(5)$_1$ WZW with $c=4$),
and the {\it Reshetikhin} model. From the latter model it is
believed that it could be described by SO(5)$_1$ WZW with $c=5/2$\cite{Alet2010}.
It turns out that this model is located near our
critical line in the phase diagram of the general spin 2 chain.
Indeed we find $c=5/2$ for the Reshetikhin model, but we leave its
connection to our critical line as an open question\cite{Alet2010}.

In conclusion, we have confirmed the
phase diagram of the AKLT--SZH chain; remarkably, also for $\beta<0.27$ a dimer
phase is present. The critical line separating the SZH and dimer phases has the same central charge $c=5/2$ and  can be described by $\rm SO(5)_1$ WZW theory.

{\it Acknowledgement.\,} We thank S. Trebst, L. Balents, Z.\,C. Gu,
S. Capponi, B.\,A. Bernevig, A. L\"auchli, and R. Thomale for
insightful discussions. We thank C.\,K. Xu for pointing out to us
the possibility of $\rm SO(5)_1$ theory. HCJ thanks D.\,N.\,Sheng
for great help on ED results. HCJ acknowledges funding from
Microsoft Station Q. SCZ is supported by the NSF under grant numbers
DMR-0904264, SR by the DFG under Grant No.\,RA\,1949/1-1, and ZYW by
NSFC and NPBR grants of MOST.


\end{document}